# Radio emissions from double RHESSI TGFs

Andrew Mezentsev[1], Nikolai Østgaard[1], Thomas Gjesteland[1,2], Kjetil Albrechtsen[1], Nikolai Lehtinen[1], Martino Marisaldi[1,3], David Smith[4], and Steven Cummer[5]

[1]Birkeland Centre for Space Science, Department of Physics and Technology, University of Bergen, Bergen, Norway, [2]Department of Engineering Sciences, University of Agder, Grimstad, Norway, [3]INAF-IASF, National Institute for Astrophysics, Bologna, Italy, [4]Department of Physics, Santa Cruz Institute for Particle Physics, University of California, Santa Cruz, California, USA, [5]Electrical and Computer Engineering Department, Duke University, Durham, North Carolina, USA

**Abstract** A detailed analysis of Reuven Ramaty High Energy Solar Spectroscopic Imager (RHESSI) terrestrial gamma ray flashes (TGFs) is performed in association with World Wide Lightning Location Network (WWLLN) sources and very low frequency (VLF) sferics recorded at Duke University. RHESSI clock offset is evaluated and found to experience changes on the 5 August 2005 and 21 October 2013, based on the analysis of TGF-WWLLN matches. The clock offsets were found for all three periods of observations with standard deviations less than 100 μs. This result opens the possibility for the precise comparative analyses of RHESSI TGFs with the other types of data (WWLLN, radio measurements, etc.) In case of multiple-peak TGFs, WWLLN detections are observed to be simultaneous with the last TGF peak for all 16 cases of multipeak RHESSI TGFs simultaneous with WWLLN sources. VLF magnetic field sferics were recorded for two of these 16 events at Duke University. These radio measurements also attribute VLF sferics to the second peak of the double TGFs, exhibiting no detectable radio emission during the first TGF peak. Possible scenarios explaining these observations are proposed. Double (multipeak) TGFs could help to distinguish between the VLF radio emission radiated by the recoil currents in the +IC leader channel and the VLF emission from the TGF producing electrons.

## 1. Introduction

The discovery of bursts of energetic photons coming out to space from the Earth's atmosphere [*Fishman et al.*, 1994], referred to as terrestrial gamma ray flashes (TGFs) stimulated significant research activity related to the connection between the TGFs and radio frequency (RF) sferics produced by different types of thunderstorm electrical discharges [*Inan et al.*, 1996; *Cummer et al.*, 2005; *Cohen et al.*, 2006; *Stanley et al.*, 2006; *Shao et al.*, 2010; *Lu et al.*, 2010; *Connaughton et al.*, 2010; *Collier et al.*, 2011; *Cummer et al.*, 2011; *Dwyer and Cummer*, 2013; *Connaughton et al.*, 2013]. Some of the current observations and modeling results attribute TGF generation to the leader development stage of the positive in-cloud (+IC) discharge, when its negative leader propagates upward between the main negative and the upper positive charge regions, transporting the negative charge upward in the cloud [*Cummer et al.*, 2005; *Stanley et al.*, 2006; *Shao et al.*, 2010; *Lu et al.*, 2010; *Celestin and Pasko*, 2011; *Celestin et al.*, 2012; *Østgaard et al.*, 2013; *Cummer et al.*, 2014, 2015]. The TGF production is suggested to occur during the leader stepping through an injection of thermal runaway electrons into the intense electric field region in front of the leader tip [*Celestin and Pasko*, 2011; *Celestin et al.*, 2012]. A competing theory involves the relativistic feedback mechanism and does not necessarily require the high field conditions present at the tip of a developing leader [*Dwyer*, 2012; *Dwyer and Cummer*, 2013]. Extensive reviews of different aspects of TGF physics are made by *Dwyer et al.* [2012] and by *Dwyer and Uman* [2014].

The possibility of the RF emissions by the TGFs themselves is currently actively discussed. The attempts to relate TGFs and radio sferics started just after the TGF discovery [*Inan et al.*, 1996]. In the series of works, TGFs were associated with the thunderstorm electromagnetic (EM) activity by use of very low frequency (VLF) sferics geolocation [e.g., *Inan et al.*, 1996; *Cummer et al.*, 2005; *Stanley et al.*, 2006; *Hazelton et al.*, 2009; *Cohen et al.*, 2006, 2010; *Connaughton et al.*, 2010; *Briggs et al.*, 2010; *Collier et al.*, 2011; *Connaughton et al.*, 2013]. *Lu et al.* [2010] analyzed one Reuven Ramaty High Energy Solar Spectroscopic Imager (RHESSI) TGF in association with the lightning mapping array mapping of the IC propagating leader and ultra low frequency (ULF) magnetic field recordings. *Cummer et al.* [2011] analyzed two Fermi Gamma-Ray Burst Monitor (GBM) TGFs that occurred close to the low-frequency (LF) magnetic field sensor which allowed to record and compare the sferic waveforms and the TGF light curves. This allowed them to relate the radio sferic emission to the TGF itself. *Connaughton et al.* [2013] presented their analysis of the Fermi GBM TGF matches with World Wide Lightning Location Network (WWLLN) sources focusing their attention on the simultaneous TGF-WWLLN matches. *Dwyer and Cummer* [2013] proposed a theory of radio emission from TGFs. *Østgaard et al.* [2013] reported simultaneous observations of a RHESSI TGF, WWLLN source, and Lightning Imaging Sensor optical signature, which allowed them to estimate the TGF production altitude and suggest that the TGF was produced by the upward propagating +IC leader. *Marshall et al.* [2013] performed a comprehensive analysis of the initial breakdown (IB) pulses of the +IC discharges and speculated that observed characteristics could be related to the TGFs, though they did not observe any related TGF for their EM data. *Cummer et al.* [2014] succeeded to measure the production altitudes of two LF sferics simultaneous with two Fermi GBM TGFs. *Cummer et al.* [2015] analyzed three Fermi GBM TGFs in associations with the LF radio recordings to define the LF sferics altitudes and relating those altitudes to simultaneously recorded TGFs during the +IC leader development.

Nevertheless, direct and unambiguous measurements of radio emissions from the TGF itself are still to be performed. Also, it is not clear yet which of the two competitive mechanisms of the TGF production (thermal runaway in the strong local transient leader field or the relativistic feedback mechanism in a weaker field), or both, or neither, takes place in the nature.

Currently, there are three working space missions providing TGF recordings: Astrorivelatore Gamma a Immagini Leggero (AGILE), Fermi GBM, and RHESSI.

The relation between Fermi GBM TGFs and WWLLN events was studied in detail by *Connaughton et al.* [2010] and *Connaughton et al.* [2013]. Fermi GBM TGFs have precise timing (down to a microsecond level) which allowed to find a significant fraction of TGFs (especially the short ones) to be simultaneous with WWLLN sources. *Connaughton et al.* [2013] interpreted their results as a manifestation of radio emissions directly from the TGF producing runaway electrons and accompanying secondary thermal electrons [*Dwyer and Cummer*, 2013].

AGILE satellite passes above the equatorial region declining from the equator by $\sim\pm 2.5°$. Recently, a new instrument configuration, which increased the TGF detection rate (due to the weaker ones) by an order of magnitude, was implemented [*Marisaldi et al.*, 2015]. With this enhanced configuration a set of TGFs simultaneous (within $\pm 200$ μs) to the WWLLN sferics has been detected. Moreover the inverse dependence of the probability of association on the TGF duration reported by *Connaughton et al.* [2013] has been confirmed.

RHESSI satellite is able to detect TGFs originating from the area $\sim\pm 46°$ of geographical latitude (due to the orbital inclination of $\sim\pm 38°$ plus the field of view up to $\sim 800$ km from the satellite footprint). Although being a very efficient instrument, RHESSI is known to have a systematic clock offset, which value (of $\sim -1.8$ ms) was not known with the appropriate accuracy [*Grefenstette et al.*, 2009]. Also, it was not clear whether this systematic clock offset has a permanent character or if it changes value with time. This offset makes any type of analysis involving precise timing comparisons between RHESSI TGFs and some other data (e.g., WWLLN detections) difficult to perform. Nevertheless, RHESSI has a very precise mission clock [*Smith et al.*, 2002] with relative accuracy level of one binary microsecond ($2^{-20} = 0.9537$ μs) and the clock drift table is being updated regularly on the RHESSI website and via the Solar Software updates (the software that is used for the processing of RHESSI data).

In the present work we analyze RHESSI TGFs together with WWLLN sources and a set of VLF and ULF magnetic field recordings performed by the receivers deployed at Duke University, Durham NC, USA. Our results open the possibility for the precise timing analysis of the RHESSI TGFs. Furthermore, we can now associate them with other types of data down to a precision level of at least better than 100 μs.

The main focus of our work is on the double- (multi) peak RHESSI TGFs that have simultaneously reported VLF radio emissions. The other three space missions capable of performing the TGF recordings also reported about detections of the double- (multi) peak TGFs.

BATSE observations revealed numerous multiple TGFs [*Fishman et al.*, 1994; *Nemiroff et al.*, 1997], which resulted from the large effective detectors' area and triggering mechanism with longer integration time [*Cohen et al.*, 2006]. Some of the double-peak events (separated by less than 1 ms) could be explained by the dead time effects [*Gjesteland et al.*, 2010]. During the BATSE experiment on board of the Compton Gamma-Ray Observatory (up to June 2000), there were no operational worldwide lightning detection networks to relate the observed multiple TGF peaks to the geolocated radio emissions. *Cohen et al.* [2006] reported one case of a three-peak TGF in association with three VLF sferics and one double-peak TGF in association with a single VLF

 **Journal of Geophysical Research: Atmospheres**  10.1002/2016JD025111sferic. However, the radio source geolocations were unknown for those TGFs, which did not allow the precise timing comparisons between the TGFs and VLF sferics.

*Dwyer et al.* [2008] reported about several BATSE TGFs with much longer durations (up to 30 ms) than typical TGFs do. One of those events had two peaks. These long events were interpreted as terrestrial electron beams that could escape the atmosphere and propagate along the geomagnetic field line. The second peak of the reported event was due to the magnetic mirroring of the electrons at the conjugate point. This type of events has a 2 orders of magnitude longer duration than typical TGFs and can be easily ruled out from our analysis.

The large fraction of Fermi GBM TGFs (∼19%) is reported to be multiples of the type A (see section 4); in addition, about 11% of the total amount of TGFs are multiples of the type B [*Foley et al.*, 2014], but only two double TGFs (one of each type) have been reported so far to match with WWLLN sferics [*Connaughton et al.*, 2010].

*Marisaldi et al.* [2014] reported that for the AGILE satellite the percentage of the multiple-peak events was about 2% of the total amount of the detected TGFs (7 out of 308), with potentially more multiples revealed by the visual inspection. The analysis for the TGF-WWLLN matches performed by *Marisaldi et al.* [2014] did not produce a positive result, probably due to the AGILE sensitivity bias toward longer (by a factor of 2.6 compared to Fermi GBM) TGFs. This negative result is consistent with conclusions of *Connaughton et al.* [2013], according to which the shorter the TGF, the more likely it has a simultaneous WWLLN detection.

In addition, *Marisaldi et al.* [2014], referring to the works of *Grefenstette et al.* [2009], *Grefenstette et al.* [2012] and T. Gjesteland (personal communication, 2013) reported that RHESSI TGFs are mostly isolated events and do not have multipeak events except for a few doubles. They conclude that "the lack of multiple-peak TGFs in the RHESSI data (compare to BATSE, Fermi-GBM, and AGILE data) is remarkable and still needs a clear explanation." Here we report that RHESSI has a fraction of multipeak TGFs of ∼3.5%, 102 out of 2779 events, that is comparable to AGILE, with potentially more multipeak TGFs with weaker peaks that we could not confidently classify as TGFs.

On 23 March 2015 the AGILE ground team has implemented a new instrument configuration, which made the instrument more sensitive to the weaker TGFs and increased the detection rate by about 1 order of magnitude [*Marisaldi et al.*, 2015]. Between 23 March and 30 June 2015 AGILE detected 288 TGFs, with three pure double-peak events revealed by the search algorithm. One double event (occurred on 6 April 2015 at 14:49:59.756416) has an associated WWLLN detection.

In the presented work we found 16 double RHESSI TGFs with simultaneous WWLLN detections and report an intriguing observational result that in all revealed cases of double- (multiple-) peak TGFs; those WWLLN detections are simultaneous with the last TGF peak. Two VLF sferics from Duke University give support to this result. We also propose possible scenarios that could explain such a behavior in the framework of the TGF generation model in the strong local leader field.

## 2. Instrumentation and Data

TGFs analyzed here were identified from the RHESSI data [*Smith et al.*, 2002] by use of the off-line search algorithm developed by *Gjesteland et al.* [2012], which is a modified version of the algorithm proposed by *Grefenstette et al.* [2009]. The search algorithm developed by *Grefenstette et al.* [2009] is focused on the cleanness of the produced TGF catalog rather than on its completeness. The algorithm of *Gjesteland et al.* [2012] detects many more weaker TGFs but requires after-search manual processing to reject different types of artifacts.

Identified TGFs were used for the search for matches with the WWLLN sources (as in *Briggs et al.* [2010], *Connaughton et al.* [2010], *Collier et al.* [2011], *Connaughton et al.* [2013], and *Marisaldi et al.* [2015]). The WWLLN catalog provides lightning geolocation and timing by the use of over 50 VLF sensors around the globe (for more information, see http://wwlln.net). The time of group arrival technique provides average accuracy of 5 km and 10 μs, which varies significantly, though with geographical origin of the storm [*Rodger et al.*, 2005, 2006; *Hutchins et al.*, 2012]. To geolocate a lightning, WWLLN needs to detect its VLF sferic at least by five stations [*Rodger et al.*, 2005].

In addition, a set of VLF and ULF magnetic field recordings performed at Duke University [*Cummer et al.*, 2005] were used for the analysis of characteristics of the radio emissions associated with TGFs. VLF and ULF sensors

MEZENTSEV ET AL.                                           RADIO FROM DOUBLE TGFS                                                                   3

at Duke University (35.975°N, 79.094°W) are the two pairs of magnetic induction coils that record horizontal magnetic fields between 50 Hz and 30 kHz (VLF) and from <0.1 Hz to 400 Hz (ULF). Timing accuracy for the VLF sensor is no worse than 20 μs and allows to attribute recorded sferics to TGFs and WWLLN sources. The ULF sensor bandwidth of 400 Hz is not suitable for timing purposes (better than 1 ms). The ULF recordings were used to unambiguously identify the polarity of the currents associated with the TGFs.

## 3. TGF-WWLLN Matches and RHESSI Clock Offsets

The performed analysis is based on the results of RHESSI TGF-WWLLN matches. *Connaughton et al.* [2013] demonstrated that a significant part of Fermi GBM TGFs (~26%) are simultaneous within ±200 μs with WWLLN sources. *Marisaldi et al.* [2015] reported a similar result on AGILE TGF-WWLLN matches with the match rate of ~14%. These results indicate that there is a significant fraction of TGFs that are simultaneous with WWLLN sources, whatever physical mechanism is responsible for such a connection.

It has been known that the RHESSI clock experiences a systematic offset of ~1.8 ms and a quasi-regular drift. The rough value of the offset was estimated by *Grefenstette et al.* [2009], with reference to the simultaneous observations of the giant flare of SGR 1806-20 on 27 December 2004 by both RHESSI and SWIFT satellites [*Palmer et al.*, 2005; *Boggs et al.*, 2007]. *Grefenstette et al.* [2009] introduced the value of 1.8 ms and commented that they assume "an uncertainty in the absolute timing of the RHESSI instrument of 1 or 2 ms."

The relative clock drift values are regularly reported in the clock drift log files which can be downloaded from the RHESSI website (this is done automatically when the Solar Software (SSW) is updated). Given that the SSW is up to date, the clock drift is automatically corrected for, but the systematic offset is never taken into account in the SSW timing procedures and the exact reference time remains unknown which makes the precise timing comparisons between RHESSI TGFs and other measurements (for instance, WWLLN detections) barely possible. The persistence of the offset value was also questionable.

To calculate the RHESSI clock systematic offset we have performed a search for TGF-WWLLN matches. In case of a systematic clock offset of a permanent character, the search for matches procedure should reveal a significant population of the TGFs matching with WWLLN detections with a certain time delay $\Delta T$. The distribution of those time delays gives the value of the clock offset and the total combined uncertainty of the TGF-WWLLN matches (WWLLN uncertainty, RHESSI positioning uncertainty, RHESSI clock uncertainty, etc.).

The TGF-WWLLN matching procedure consisted of two stages: the coarse search and the fine search. The coarse matches were collected for the further fine search if the WWLLN source appeared to be in a ±5 ms time window around the TGF trigger time and within a circle of a radius of 800 km from the RHESSI subsatellite point on the Earth surface. This coarse search procedure found 397 rough TGF-WWLLN matches out of the total amount of 2779 TGFs revealed by the *Gjesteland et al.* [2012] search algorithm during the observation period from June 2002 to May 2015.

The light curves of the 397 TGFs selected after the coarse search for rough matches with WWLLN sources binned into the 50 μs time bins were fitted by the Gaussian fit to find out the peak times (as a peak of the Gaussian fit) of these TGFs. After that the fine search for TGF-WWLLN matches was performed to produce the distribution of the time differences $\Delta T = t_{TGF} - t_{WWLLN}$ between the TGF peak times and the WWLLN source times. TGF peak times were corrected for the propagation time from the WWLLN sources to the RHESSI satellite, the altitude of the sources was assumed to be 15 km in all cases [*Østgaard et al.*, 2013; *Cummer et al.*, 2014, 2015; *Marisaldi et al.*, 2015].

Figure 1a shows the resulting time differences $\Delta T$ between the TGF peak times and WWLLN source times plotted against the occurrence dates, demonstrating how the $\Delta T$ values change with time, year to year. Almost all of the TGFs after the coarse search procedure clustered around the three distinct offset values.

The whole observation period is then divided into the three time intervals. The first interval starts from the beginning of the analysis period and lasts up to 5 August 2005. On that date an update of the RHESSI timing procedure was implemented by the RHESSI ground team. This resulted in increased accuracy of the reference time stamp evaluation procedure and introduced a shift in the clock offset by a value of ~500 μs. Also, due to the poorer accuracy of the time referencing during this first observational period, the standard deviation $\sigma_1$ has a larger value compared to the later periods when the accuracy of the time reference evaluation procedure was increased. Figure 1b shows the 20 μs binned distribution of $\Delta T$ for this period fitted by the Gaussian fit with mean and standard deviation values corresponding to the Gaussian fit parameters.

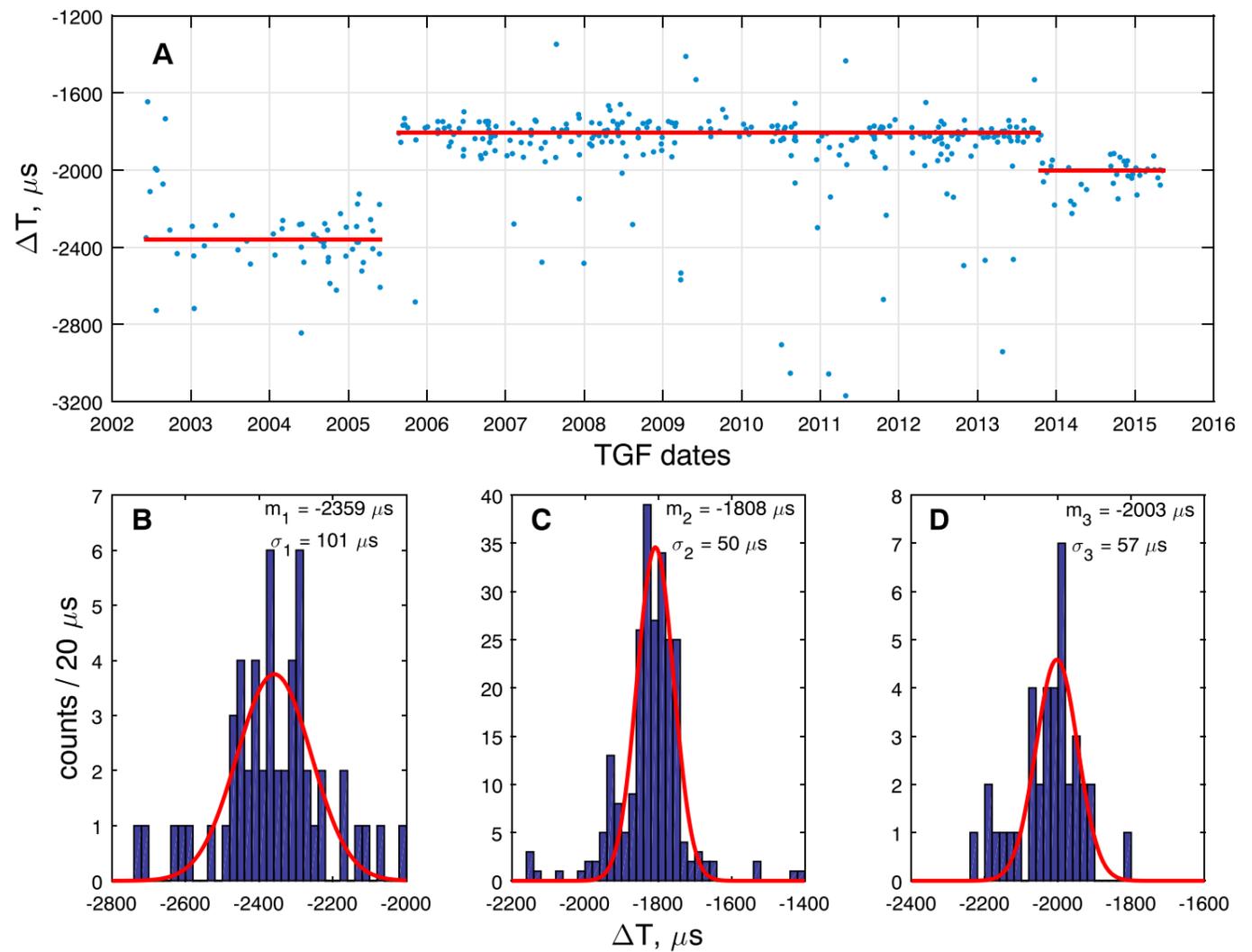

**Figure 1.** Distribution of the time difference ΔT between the TGF peak times and the WWLLN times. (a) ΔT versus TGF date. Three horizontal red lines show the mean values of the three clock offsets. (b–d) Histograms and their Gaussian fits of the ΔT distributions for the three observational periods.

The second, the longest period with a constant clock offset started after 5 August 2005 and finished 21 October 2013 (Figure 1c), when, again, a certain update of some data processing procedures was implemented by the RHESSI ground team, which, in turn, affected the RHESSI reference time stamp procedure, introducing a constant delay of ~200 μs. The last period started 21 October 2013 and goes up to the current moment (Figure 1d).

In Figure 1a all three offset values are shown as thick red horizontal lines. The results on the RHESSI clock offset values are summarized in Table 1. Notice that the offset mean values and their standard deviations for the three observation periods could be calculated in different ways (giving similar results, though). For the consistency we use the mean value $m$ and the standard deviation $\sigma$ given by the Gaussian fit of the ΔT distributions.

Subtracting the obtained values of the offsets from the time differences ΔT between the TGF peak times and WWLLN detection times and keeping only those TGFs with $|\Delta T| \leq 400$ μs (corrected for the clock offsets), we produce the distribution for the set of the 335 TGF-WWLLN matching pairs. The histogram for this

**Table 1.** RHESSI Clock Offset Values[a]

| Observation Period | Offset Value $m$ (μs) | Standard Deviation $\sigma$ (μs) |
| --- | --- | --- |
| 1 June 2002 to 5 August 2005 | −2359 | 101 |
| 5 August 2005 to 21 October 2013 | −1808 | 50 |
| 21 October 2013 to 31 May 2015 | −2003 | 57 |

[a]Gaussian fit parameters of the ΔT distributions for the three observation periods.

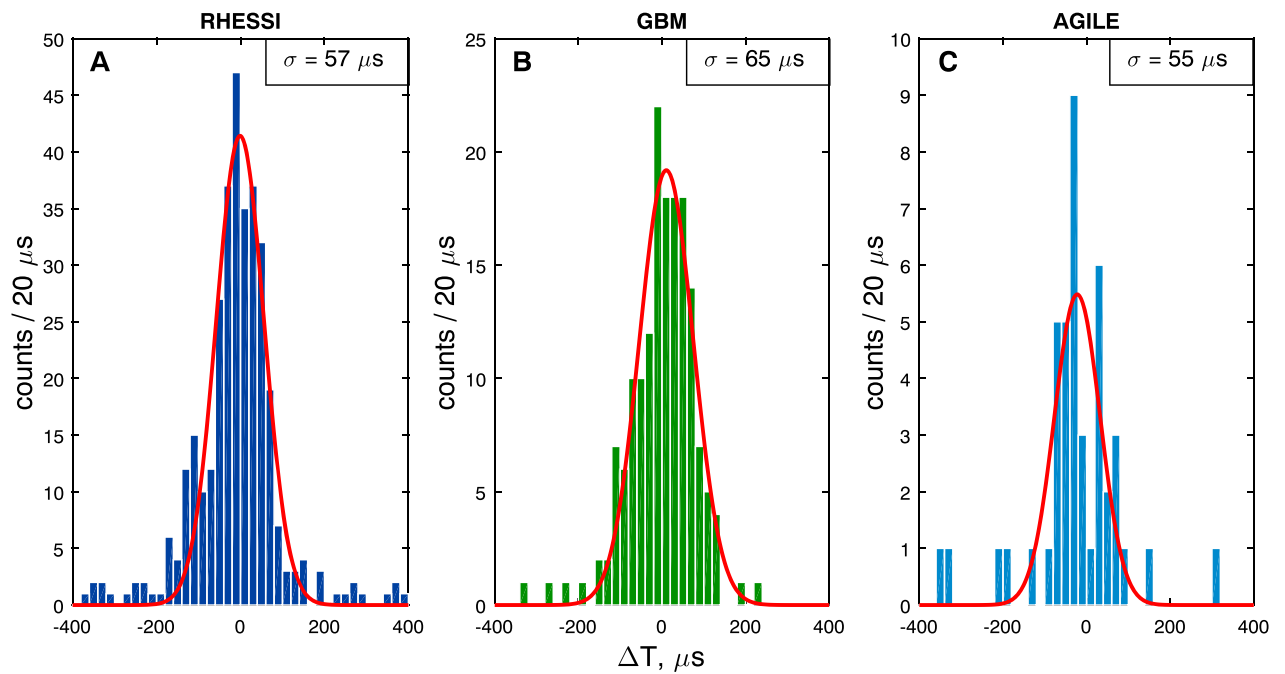

**Figure 2.** Comparisons of the RHESSI, Fermi-GBM, and AGILE TGF matches with WWLLN detections. (a) Distribution of the time differences $|\Delta T|$ between RHESSI TGF peak times and WWLLN detections. TGF peak times are corrected for the RHESSI clock offsets. Gaussian fit standard deviation $\sigma$ is equal to 57 μs. (b) Distribution of the time differences $|\Delta T|$ between Fermi TGF peak times and WWLLN detections. Gaussian fit standard deviation $\sigma$ is equal to 65 μs. Adopted from *Connaughton et al.* [2013]. (c) Distribution of the time differences $|\Delta T|$ between AGILE TGF peak times and WWLLN source times. Gaussian fit standard deviation $\sigma$ is equal to 55 μs. Adopted from *Marisaldi et al.* [2015]. For all three distributions the bin size is equal to 20 μs, TGF peak times are corrected for the propagation time from the WWLLN source location to the satellites; the source altitude is assumed to be 15 km for AGILE and RHESSI TGFs.

distribution with the bin size of 20 μs is given in Figure 2a. Following the criterion for simultaneity between the TGFs and WWLLN sources, introduced by *Connaughton et al.* [2013], we call a TGF and a WWLLN source simultaneous if the time difference between the TGF peak time and the WWLLN source time is $|\Delta T| = |t_{TGF} - t_{WWLLN}| \leq 200$ μs. Within the analysis period we found 314 simultaneous TGF-WWLLN pairs from the total amount of 2779 TGFs detected during this period.

For the comparison purpose we reproduce corresponding results obtained by *Connaughton et al.* [2013] and *Marisaldi et al.* [2015] in Figures 2b and 2c, correspondingly. In all three cases the results are binned into 20 μs bins and fitted by a Gaussian fit. Standard deviations are remarkably close for all three distributions. This allows to assume that the main contribution into the uncertainty of the TGF-WWLLN matches is given by the WWLLN uncertainty and by the natural variability of the process itself.

Interestingly, in addition to the confirmation by the RHESSI ground team about the changing dates, our result was independently validated by the results of analysis by using the method of *Østgaard et al.* [2015] on weak TGF search. Their work was based on stacking the RHESSI light curves associated with all WWLLN detections within the RHESSI field of view of 800 km radius, corrected for the WWLLN source-satellite travel time.

The method used in *Østgaard et al.* [2015] shows that there is a population of weak TGFs that cannot be detected by the existing search algorithms, because each individual weak TGF cannot be distinguished from the background, but hundreds of weak TGFs would give a significant signal exceeding the background level. Figure 3 shows the stacked RHESSI data for the three observation time periods specified above and associated with different RHESSI clock offsets. Figure 3a corresponds to the first observation period when RHESSI reference time was generated with lower precision and the WWLLN network had a lower lightning detection rate and accuracy [*Rodger et al.*, 2006]. These factors result in a smaller and wider peak above the mean background level, but still, it is discernible. Figure 3b corresponds to the longest second period and gives a huge peak. Figure 3c shows the peak for the last observation period. Red vertical dashed lines show the RHESSI clock offsets for each period and serve for the guidance purpose. The shift between the peaks of the second and third periods is clearly seen from Figure 3 and closely corresponds to 200 μs obtained by our analysis on the TGF-WWLLN matches.

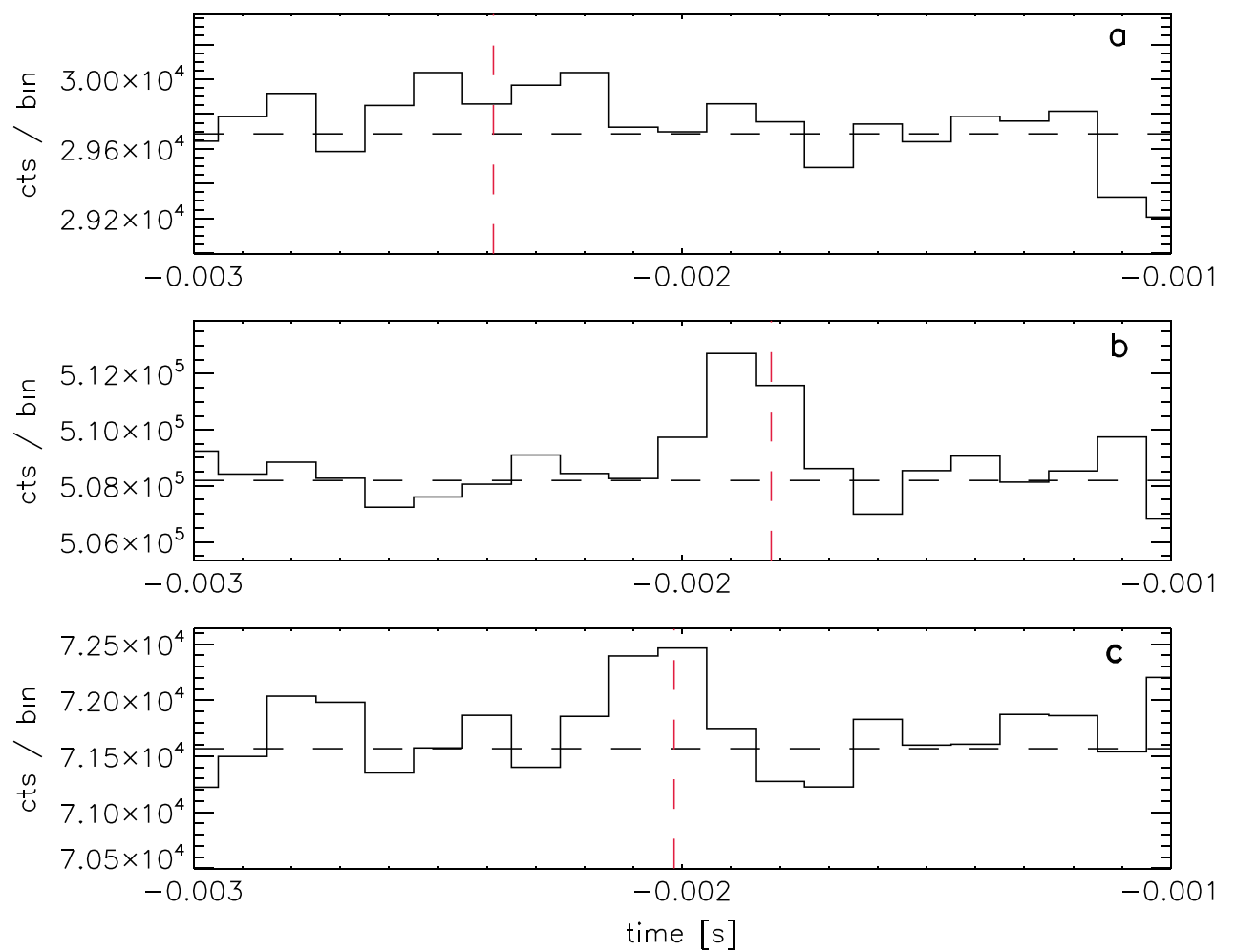

**Figure 3.** Stacked RHESSI data simultaneous with all WWLLN detections in the RHESSI field of view, binned at 100 μs. (a) Data collected from August 2004 to August 2005. (b) RHESSI data between August 2005 and October 2013. (c) Data from October 2013 to December 2014. Black horizontal dashed lines show the mean background level. Red vertical dashed lines show the values of the RHESSI clock offsets calculated by the TGF-WWLLN match analysis.

The main result of this analysis is that the set of RHESSI systematic clock offsets were measured with a better than 100 μs uncertainty for the whole period of RHESSI-WWLLN observations. These results now open the way for all research groups to use RHESSI TGFs for comparative analyses with other types of measurements with the unprecedented level of timing accuracy of at least 100 μs. Comparison between our result and results reported by *Connaughton et al.* [2013] and *Marisaldi et al.* [2015] reveals remarkably close values of the standard deviations of $\Delta T$ distributions for the RHESSI, Fermi GBM, and AGILE satellites. This allows to infer that the main contribution into the overall uncertainty of the $\Delta T$ distributions in Figure 2 is given not by the satellites' clocks (which have a relative precision down to a microsecond level for Fermi GBM, AGILE, and RHESSI) but by the WWLLN uncertainty and the natural variability of the TGF-related radio emission process itself. This last circumstance also might shed some light onto the mechanisms of the TGF generation.

## 4. Radio Emission Associated With Double TGFs
### 4.1. RHESSI TGFs
Within the analysis period (from June 2002 to May 2015) we found 16 multiple (15 double and 1 four) peak TGFs with simultaneous WWLLN detections. For two of those 16 events we also have VLF and ULF radio waveforms recorded at Duke University, which allow us to analyze the time evolution of the radio emissions in addition to the WWLLN data. We distinguish the two types of double (multiple) TGFs. Double (multiple) TGFs of type A consist of two (or more) clearly distinct TGF peaks, separated by a time interval of 1.0 to 10.0 ms. The multipeak TGFs of type B consist of two (or more) TGF peaks with peak separation intervals shorter than 0.5 ms, which leads to the overlapping TGF pulses. This type of TGFs could be interpreted as a single long TGF with subpulses in it. Table 2 summarizes the characteristics of the 16 found multiple TGFs simultaneous with WWLLN sources.

**Table 2.** Double-Peak RHESSI TGFs With WWLLN Matches

| Event Date, | RHESSI Trigger Time / WWLLN Time | RHESSI Position, (lat, lon, H), (°N, °E, km) / WWLLN Source Location, (lat, lon), (°N, °E) |
|---|---|---|
| a. 29 May 2005[b] | 21:12:27.844 | (19.1022, 100.7447, 569.1246) |
| | 21:12:27.844813 | (19.6311, 101.2218) |
| b. 17 Sep 2006[b] | 09:19:34.802 | (12.9793, −78.0183, 570.6586) |
| | 09:19:34.802356 | (10.8495, −76.8721) |
| c. 16 Oct 2006[a] | 07:21:17.070 | (35.3910, 16.5117, 568.1331) |
| | 07:21:17.072720 | (35.4021, 17.3307) |
| d. 2 Feb 2007[b] | 04:56:25.415 | (−4.3058, 149.0278, 574.3248) |
| | 04:56:25.415399 | (−6.0929, 150.1318) |
| e. 4 Jul 2007[a] | 11:57:43.768 | (−3.5945, 82.2431, 559.5059) |
| | 11:57:43.769547 | (−2.8432, 83.1585) |
| f. 13 Aug 2008[b] | 03:19:41.485 | (16.9395, 157.5847, 547.2706) |
| | 03:19:41.485913 | (14.8887, 156.8908) |
| g. 18 Nov 2008[b] | 14:22:17.990 | (1.0332, 24.0291, 555.9374) |
| | 14:22:17.989975 | (−2.2717, 21.8266) |
| h. 24 Dec 2010[b] | 14:30:03.478 | (-22.5388, 33.3270, 556.4579) |
| | 14:30:03.476223 | (−23.0724, 32.6479) |
| i. 31 May 2011[a,r] | 08:17:56.852 | (9.6249, −73.2309, 548.2862) |
| | 08:17:56.852317 | (6.9742, −74.0174) |
| j. 4 Nov 2011[b,r] | 07:54:52.210 | (8.7851, −64.3252, 560.1132) |
| | 07:54:52.209849 | (6.9439, −63.8950) |
| k. 4 Mar 2012[b] | 03:17:19.496 | (−21.5790, −146.0627, 557.9726) |
| | 03:17:19.496495 | (−19.4167, −146.5719) |
| l. 11 Aug 2012[a] | 19:38:23.899 | (2.0207, 151.8344, 547.5548) |
| | 19:38:23.898857 | (4.2309, 151.6465) |
| m. 11 Jun 2013[a] | 06:24:42.417 | (19.2280, −91.9027, 538.6153) |
| | 06:24:42.417550 | (17.9277, −93.9624) |
| n. 22 Sep 2014[a] | 15:26:39.765 | (13.5863, −98.0952, 518.4464) |
| | 15:26:39.766795 | (15.1632, −99.0264) |
| o. 20 Nov 2014[a,m] | 17:58:58.402 | (2.7147, 13.4215, 505.8788) |
| | 17:58:58.403513 | (2.4024, 13.5195) |
| p. 8 May 2015[a] | 18:19:17.263 | (3.0615, 78.6825, 505.6521) |
| | 18:19:17.264871 | (3.1914, 76.7104) |

[a] A type double TGF.
[b] B type double TGF.
[r] Radio waveforms present.
[m] Multiple (four) peak TGF.

Figure 4 shows all 16 TGF light curves with superimposed WWLLN marks (dashed red vertical lines) corrected for the source to satellite propagation time and for the RHESSI clock offsets. There are eight double TGFs of type A with two well-separated TGF peaks, and seven double TGFs of type B with two closely located but clearly distinct TGF pulses. One event consists of four distinct TGF peaks. In all 16 cases the WWLLN detection is simultaneous with the last peak of the multipeak TGFs.

One case of a double TGF (see Figure 4c) has two associated WWLLN sources: the first was detected ∼700 μs prior to the first TGF peak and the second WWLLN source was simultaneous with the second TGF peak. *Gjesteland et al.* [2015] discussed this particular TGF to a great level of detail and argued that the first WWLLN source did not relate to the first TGF peak. From our analysis we also conclude that having the timing accuracy of the RHESSI clock better than 100 μs, WWLLN accuracy better than 40 μs, the first WWLLN detection occurs ~700 μs prior to the first TGF peak, which indicates that they are not related. The second WWLLN source is simultaneous with the second TGF peak, which is consistent with the observed tendency of having WWLLN detections simultaneous with the last TGF peak in multipeak TGFs.

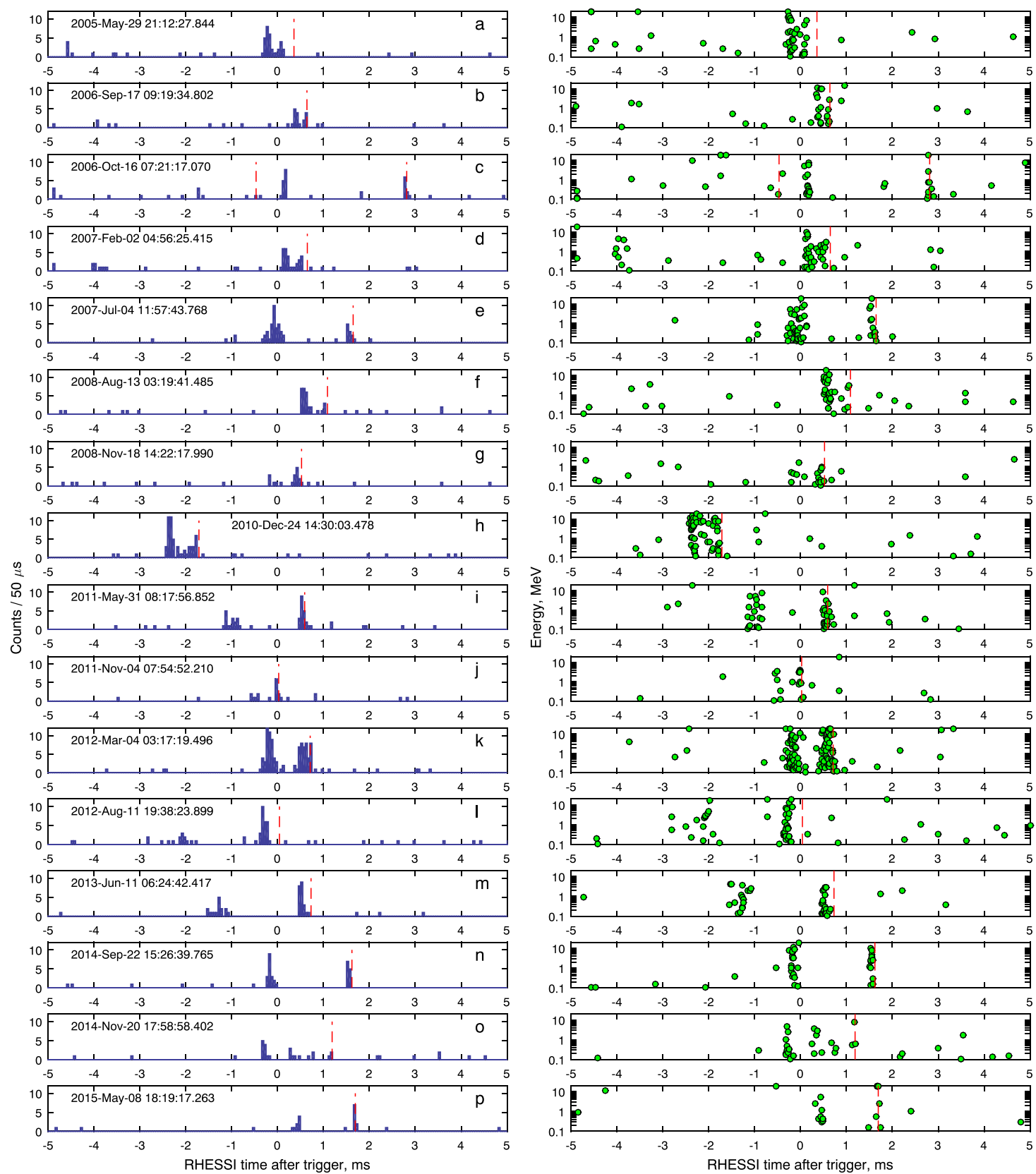

**Figure 4.** (a–p) Sixteen multipeak TGFs with simultaneous WWLLN detections (red dashed lines). Time axes are centered on the TGF trigger times (RHESSI time) highlighted for each event. (left column) TGF light curves showing (>100 keV) photon counts per 50 μs. (right column) Photon energies versus time. WWLLN time is corrected for the light travel time between the WWLLN source and RHESSI and accounted for the RHESSI clock offset.

A remarkable example of a double-peak TGF occurred on 4 March 2012 (see Figure 4k). In this example two distinct well separated by ~800 μs TGF peaks are almost identical to each other in terms of the photon counts, photon energies, and TGF durations. WWLLN detection is simultaneous to the second TGF peak, while one might expect two WWLLN detections, simultaneous to each of the two identical TGF peaks, if the radio emission responsible for WWLLN detection was caused by the TGF itself.

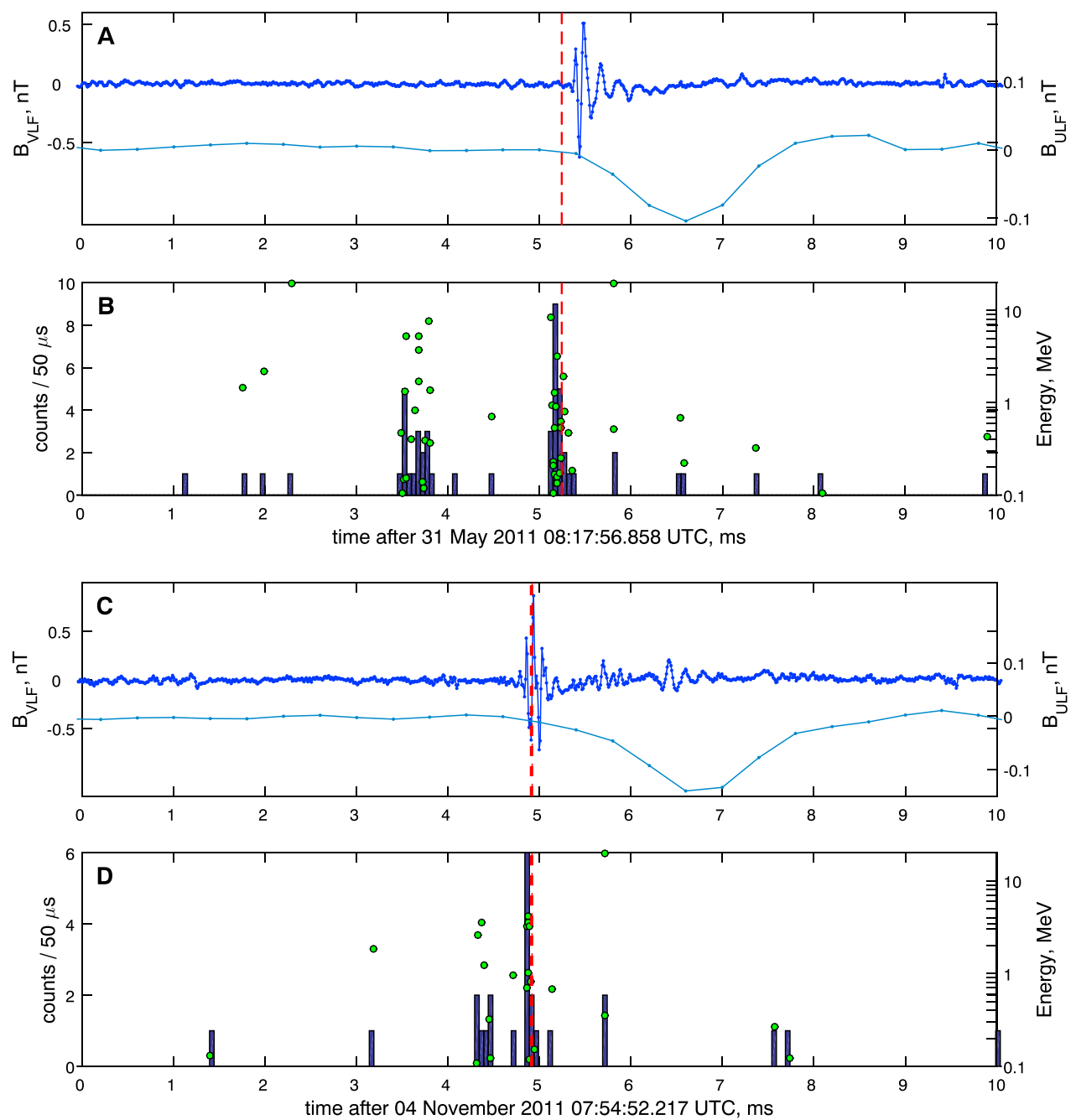

**Figure 5.** Two double TGFs with simultaneous WWLLN detections (red dashed vertical lines), VLF, and ULF waveforms recorded at Duke: (a, b) 31 May 2011 and (c, d) 4 November 2011. (Figures 5a and 5c) VLF (left axes) and ULF (slowly varying curves, right axes) radio recordings. (Figures 5b and 5d) RHESSI photon counts and energies versus time. See text for details.

Analysis of the WWLLN activity around this flash on 4 March 2012 at 03:17:19.496 UTC, (19.4167°S, 146.5719°W) reveals that this WWLLN detection was the first one in a sequence of four WWLLN sources originated from the same location of ~10 km radius (which is within WWLLN uncertainty). Second detection came 34 ms after the first one, the last two occurred ~360 ms, and ~510 ms after the first detection. These four detections presumably belonged to the same complex flash and were isolated from the other flashes detected by WWLLN in that region by 8 and 13 s before and after the flash. According to *van der Velde et al.* [2006], *Lu et al.* [2013], and *van der Velde et al.* [2014], this scenario (with the sequence of VLF pulses) is very common for the combined flashes which start as an upward propagating vertical +IC (could be bidirectional), then develop horizontally and end up as a series of cloud-to-ground (CG) discharges (the last two or three WWLLN detections in our case). The TGF is supposed to be generated during the vertical +IC leader progression.

On the other hand, considering this specific event, we have to admit that the lack of the radio recordings does not allow to rule out the possibility of radio emission simultaneous with the first TGF peak, which for some reason was not resolved by the WWLLN; e.g., because its separation algorithm might not be capable to separate the first sferic from the second one or the first sferic was detected by less than five WWLLN stations. This circumstance shows the necessity of involving the radio sferic recordings into the analysis in addition to WWLLN data which provide the source location but cannot give the temporal evolution of radio emissions.

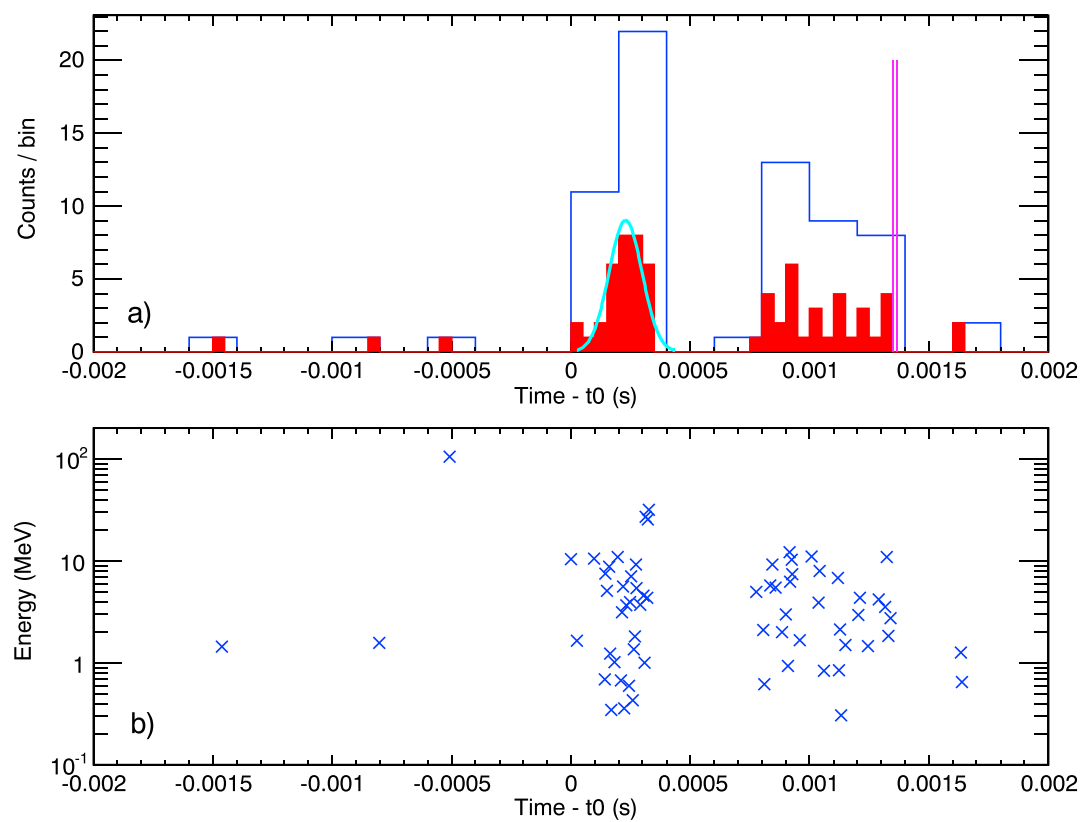

**Figure 6.** Double AGILE TGF with WWLLN match. (a) Red histogram has 50 μs bins, blue histogram has 200 μs bins. Cyan line shows the best Gaussian fit for the first TGF. The second TGF looks like a superposition of several short weak pulses. Magenta line shows the WWLLN detection (corrected for the light travel time). (b) Photon energy versus time.

Figure 5 shows two double-peak TGFs matched with WWLLN, together with the Duke radio recordings of VLF and ULF waveforms. The first TGF is a double TGF of type A with two peaks separated by ∼1.7 ms, occurred on 31 May 2011 (see Figure 4i and Table 2 for details). Figure 5a shows the VLF and ULF radio waveforms recorded at Duke University, associated with this TGF. The upper blue curve corresponds to the azimuthal VLF magnetic field component $B_{VLF, \varphi}$, while the lower blue curve shows the azimuthal ULF magnetic field component $B_{ULF, \varphi}$, relative to the Duke-WWLLN source direction. Radial magnetic field components are not shown, because they are small and do not contribute to the sferic, which supports the idea that the observed sferic arrives from the WWLLN source direction. Figure 5b contains the TGF (>100 keV) photon counts per 50 μs together with the individual photon energies versus time. The WWLLN detection is shown as the red vertical dashed line in both panels. All times (RHESSI, WWLLN) are recalculated to the Duke reference time. Figures 5c and 5d show the analogous measurements for the double TGF of type B that occurred on 4 November 2011 (Figure 4j).

The distances between the WWLLN sources and the Duke receivers were equal to 3253 km and 3569 km, which is reasonably close for the VLF and ULF sferics detection. The analysis of the WWLLN activity around Duke University (in a radius of 10,000 km) combined with the ratio between the north-south and east-west magnetic field components allows the unambiguous attribution of these two WWLLN detections to the two observed VLF radio sferics. In Figures 5a and 5c both ULF pulses show the upward transfer of negative electric charge, which is consistent with a +IC leader current or with the TGF producing RREA and its secondary electrons accelerated upward.

The timing comparison between the two TGFs in Figure 5 with the VLF radio recordings shows that the radio emission can only be related to the second TGF peak in both cases, exhibiting no detectable signal above the noise level during the first peak of these double TGFs. In both cases the amplitude of the VLF sferic is larger by a factor of 20 compared to the amplitude of the VLF signal during the first TGF peak.

These two examples are important for our understanding of the physical processes. The WWLLN produces its locations based on the TOGA (time of group arrival) processing [*Rodger et al.*, 2006], where the TOGA itself is produced at the detection site by the use of a sophisticated algorithm and then sent to the main processing site. The waveforms are not recorded, so the time evolution of the radio emission cannot be examined. Therefore, one might put forward an argument that the absence of a WWLLN detection does not necessarily mean the absence of a real radio emission. This argument makes the independent radio recordings so important for our analysis.

The brief summary given below describes the published results on the multipeak TGF-radio associations for the other three space missions capable of performing the TGF recordings.

### 4.2. BATSE, Fermi GBM, and AGILE TGFs

One three-peak TGF and one double-peak TGF were detected by BATSE in association with VLF radio sferics [*Cohen et al.*, 2006]. However, the lack of geolocations of the radio sources could not allow the precise timing analysis between the TGFs and VLF sferics.

There are two double-peak Fermi GBM TGFs reported to be simultaneous with WWLLN sferics [*Connaughton et al.*, 2010]. WWLLN sferics were simultaneous with the last (second) TGF peak for both of these double TGFs [see *Connaughton et al.*, 2010, Figure 1], which is consistent with our findings for the RHESSI TGFs.

There is one known reported multi- (three-) peak TGF detected by AGILE before the new instrument configuration has been implemented on 23 March 2015, which has a simultaneous WWLLN detection [*Marisaldi et al.*, 2014]. Remarkably, the WWLLN source is simultaneous with the last (third) TGF peak [*Marisaldi et al.*, 2014], which is consistent with our present results. After 23 March 2015 one more multipeak TGF with simultaneous WWLLN detection was found. The inspection of the light curve shown in Figure 6 revealed that the TGF consists of the first strong TGF peak and the second one which looks like a superposition of several weaker peaks with the WWLLN detection simultaneous to the last weaker peak in the sequence. This AGILE TGF is in agreement with our results for the RHESSI TGFs. No contradicting examples were found in the AGILE data up to now.

## 5. Discussion

The results of the multipeak TGF observations associated with WWLLN detections show that in all 16 cases revealed for RHESSI TGFs the WWLLN detection is found to be simultaneous with the last TGF peak. This asymmetry pattern in TGF-WWLLN relation might reflect some unknown aspects of the TGF generation process and needs its explanation. Also such a behavior possibly refers to certain peculiarities in TGF-related radio waves generation. In this section we discuss several possible scenarios that could explain our observations, though we are not pretending to propose a complete picture of the process.

For the quantitative estimate of the probability of the observed results we consider a simplified model where each TGF has a simultaneous VLF sferic which can be detected by the WWLLN with a detection rate $p \in (0; 1)$. Then the probability $P$ of WWLLN detection of only the last TGF peak in $N$ double TGFs is given by

$$P(p; N) = (1-p)^N p^N. \qquad (1)$$

This probability function for any fixed number of double TGFs $N$ reaches its maximum value for the detection rate $p = 1/2$. For 16 double TGFs this probability will not exceed the value of $P(0.5; 16) \approx 2.33 \cdot 10^{-10}$. If we consider only those double TGFs of type A which consist of clearly separated peaks, so that the VLF sferics will not overlap and cause difficulties for WWLLN to separate them, then eight double TGFs of type A will give the probability not exceeding the value of $P(0.5; 8) \approx 1.53 \cdot 10^{-5}$. These simplistic estimations show that the discussed phenomenon is unlikely to be of a probabilistic nature and certain physical causes drive this multipeak TGF-WWLLN asymmetry. Note that here we considered only RHESSI TGFs, and the inclusion of the two reported Fermi GBM and two found AGILE double- (multiple-) peak TGFs simultaneous with WWLLN detections makes it even less likely that the observed phenomenon is a result of a coincidence.

A simple explanation of the observed pattern is that WWLLN has a tendency to detect the last sferic in a sequence of closely spaced peaks, probably due to peculiarities of the sferic separation algorithm used by WWLLN. However, the probability estimate for eight double TGFs with well-separated peaks combined with the two cases of simultaneous VLF radio recordings (Figure 5) makes this explanation unlikely.

One of the possible scenarios of the TGF-VLF coupling is that the electrons producing TGFs emit VLF radio waves too weak to be detected on ground, either because of the intrinsic weakness of the emission or because the emission bandwidth lays in a different (higher) frequency range. In this scenario TGF production is firmly linked to the +IC leader development [*Celestin and Pasko*, 2011; *Celestin et al.*, 2012], to what is called initial breakdown (IB) [*Marshall et al.*, 2013]. During a +IC flash the gap between the main negative charge center (MNCC) and the upper positive charge center (UPCC) is spanned by the upward negative stepped leader during a process called initial breakdown [*Coleman et al.*, 2003; *Winn et al.*, 2011; *Marshall et al.*, 2013]. The initial conducting stem of the leader emerges due to a poorly understood process called "fast positive breakdown" [*Rison et al.*, 2016] between the two charge layers closer to the MNCC. After that the leader develops upward [*van der Velde et al.*, 2006] producing IB pulses in VLF-LF frequency range [*Marshall et al.*, 2013] when stepping. Growing leader concentrates huge potential drop (proportional to leader length and total potential difference between the MNCC and the UPCC) in front of its tip together with the local electric field enhancement, providing necessary conditions for generating TGFs. The most favorable conditions for the TGF production occur before the "attachment," when the maximal potential drop and the highest local electric field are concentrated in front of the leader tip. By the "attachment" we mean the connection between the upward growing negative leader and downward growing positive counterleader developing from the UPCC. The appearance of the counterleader seems to be plausible, though at the moment we do not have measurements that would support this idea. This is our working hypothesis. VLF sferic emitted by the recoil current after the attachment is expected to be more powerful compared to sferics emitted during stepping, simply because the channel length and the accumulated charge on it are maximal before the attachment.

This scenario does not contradict the observed asymmetry in TGF-WWLLN-VLF association. Indeed, a multiple TGF can be generated during the vertical +IC leader progression, with most favorable conditions for the last TGF peak to occur just before the leader attachment. After the attachment the most powerful VLF sferic is emitted by the recoil current wave running through the established leader channel (weaker sferics (IB pulses) can also be emitted during the stepping, rebrightening shorter segments of the channel). Thus, the most favorable conditions for the TGF production just before the attachment and the most powerful VLF sferic emitted just after the attachment might result in the observed asymmetry in multipeak TGF-WWLLN-VLF association.

Another argument supporting this scenario is that the ULF pulse is present in both examples of the simultaneous double TGF-WWLLN-VLF observations shown in Figures 5a and 5c. The presence of the accompanying ULF pulses in most cases of TGF-related VLF emission was reported in *Lu et al.* [2011]. Those ULF pulses refer to slow vertical currents transferring the negative charge upward. Within the considered scenario both VLF and ULF emissions have their natural interpretation as the recoil current (analogous to a return stroke (RS) in a cloud-to-ground (CG) discharge) and the slower current (analogous to a CG continuing current) through the established channel after the recoil current wave.

Another possible scenario consistent with our observations could involve the azimuthal asymmetry of the radio emission, when this emission is generated by a nonvertical current element. Here we propose a concept of such a scenario. In the case of vertical antenna its emission directional diagram is azimuthally symmetric (here we consider the antenna's properties only and do not account for the asymmetry caused by the propagation effects in the Earth-ionosphere waveguide). The nonvertical current element has an azimuthally anisotropic directional diagram, emitting its maximal power along the main lobe of the directional diagram. This results in a situation when those WWLLN receivers located along the main lobe would be able to detect the emitted sferic and those located along the less favorable directions will not detect it.

The vertical tilt of the TGF axis could be caused by the tortuosity of the propagating +IC leader, which does not propagate straight upward and experiences random deviations from the vertical direction. The leader brings its own strong local field in front of its tip, which is oriented along the axis of the leader tip segment. If a TGF is generated in such a vertically tilted electric field, then the runaway and slow secondary electrons responsible for the radio emission [*Dwyer and Cummer*, 2013] experience acceleration along this nonvertical axis, which makes the directional diagram of such an emitting antenna azimuthally anisotropic and results in lower detectability of the emitted sferics by WWLLN.

Figure 7 illustrates the proposed scenario. General view and geometry is shown in Figure 7a. Three black rods on the ground stand for the three WWLLN receivers (though, WWLLN needs at least five stations detection to geolocate a sferic). Thundercloud has a typical charge structure. Negative leader of the +IC flash propagates from the main negative charge region to the upper positive charge region upward. Here we consider only the vertical stepped leader propagation stage of the IC flash, because the later stage of the IC flash with the extensively growing horizontal leaders inside the MNCC and the UPCC has never been reported to be related to the TGF generation. The first TGF occurs in the middle of the leader propagation path, when the leader tip bends from the vertical axis. The second TGF occurs at the final stage of the leader propagation, when the leader tip is oriented vertically. Two TGFs are shown as colored (magenta and cyan) cones with half angle of ∼40° [*Gjesteland et al.*, 2011] with axes highlighted by the dashed lines of the same colors. The RHESSI satellite is passing by the overlapping area of the two cones, so that it detects both TGFs (Figure 7b, top histogram). Radio emissions from the TGFs are shown as colored (magenta and cyan) arcs, illustrating that radio emission from the second (vertical) TGF is azimuthally isotropic and from the first one (nonvertical) is azimuthally anisotropic. Conjectural sferics from three WWLLN receivers are shown in Figure 7b, with a WWLLN detection, associated with the second TGF marked as a dashed vertical line. Azimuthally isotropic radiation produces waveforms at all three receivers. The first TGF, because of its emission anisotropy, produces a waveform at one receiver only, so that WWLLN could not evaluate the source location of this event.

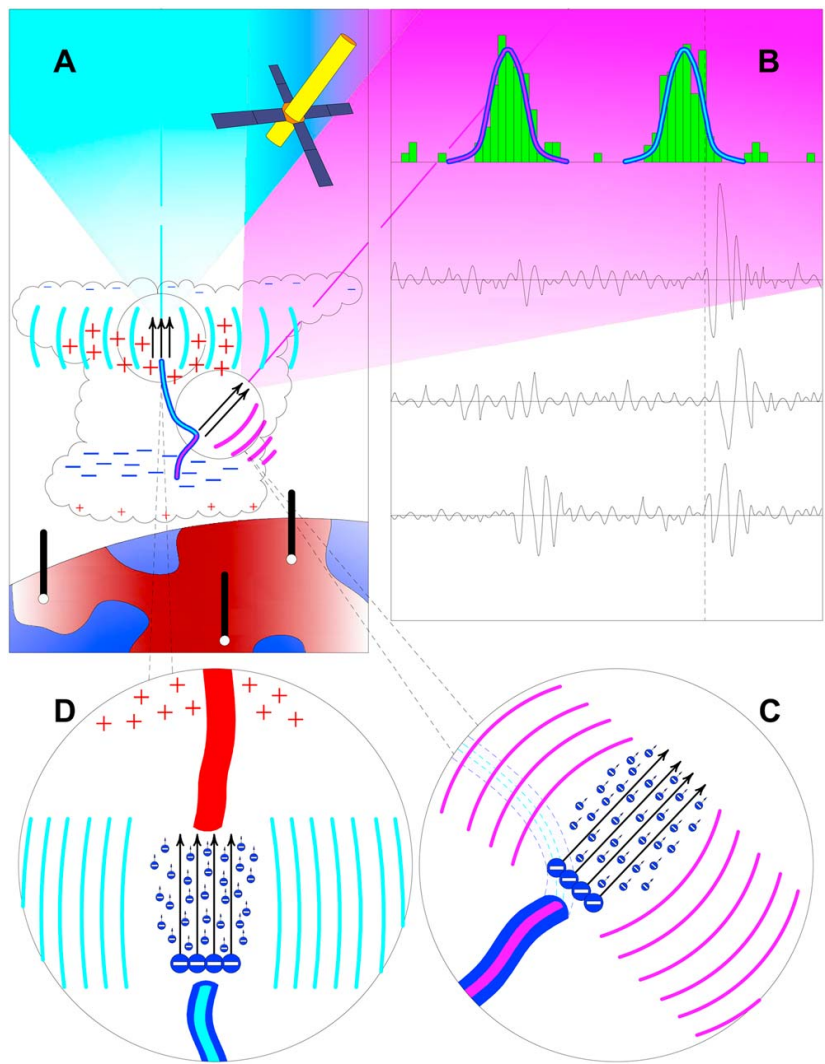

**Figure 7.** Scenario of a double TGF production by a tortuous leader. See text for the details.

The insets, Figures 7c and 7d, show the zoomed view on the two TGF production occurrences. The big blue circles with long black arrows indicate the RREA electrons, while the small blue circles with short arrows correspond to the secondary low-energy electrons produced by the RREAs, which are supposed to be responsible for the current pulses producing the radio emissions (shown by light blue and purple arcs). Strong local electric field of the leader tip that accelerates the RREA is directed along the leader tip axes in both cases. The thick vertical red line in Figure 7d represents the growing positive counterleader before the attachment.

Figure 7 illustrates the qualitative picture of the tilted leader scenario, whereas some quantitative bounds can be obtained from the real VLF recordings given in Figures 5a and 5c. The amplitude of the VLF sferics simultaneous with the second TGF peak in both cases is by a factor of 20 larger than the signal level during the occurrence of the first TGF peak. Let us assume that in both cases the TGF peaks were associated with similar leader extensions, where the first extension was tilted and the second one was essentially vertical. The tilted leader extensions can be decomposed into horizontal and vertical components, since the radiation processes are linear in this frequency range. In order for the radiation from the tilted extensions to be by a factor of 20 smaller than the radiation from the vertical ones, these tilted segments have to be no more than $\arcsin(1/20) \approx 3°$ away from the horizontal plane. For RHESSI to be in the beam of two TGFs that are beamed 87° apart from each other, one vertical, one almost horizontal, would be an extraordinary coincidence.

Thus, the two events shown in Figure 5 would require an extremely lucky observing geometry to be consistent with the proposed idea of the tilted leader. Therefore, this scenario merits more quantitative testing when the ratio of VLF radiation from multipeak TGFs can be better measured or bounded.

Another possible counterargument against the proposed scenario could be that there is not any specific reason for the upper part of the leader channel to be more vertical than its lower segments. This makes it unclear why the last TGF is always the one oriented vertically, while the earlier TGFs have tilted axes. However, after the attachment the recoil current wave along the whole leader channel inevitably generates VLF sferic, because (contrary to the individual leader steps) the horizontal displacement of the whole leader channel is much shorter compared to its vertical extent. This makes it effectively a vertical current element, producing VLF radio emission detectable by the WWLLN-like systems from the ground. In this case the last TGF might be also tilted from vertical and its VLF emission will be masked by the VLF sferic generated by the recoil current wave running through the leader channel.

The last argument gives motivation to focus attention and efforts on the search and analysis of radio emissions simultaneous with the first TGF peak in multipeak TGFs, because those emissions more likely will not be masked by radio emissions from leader currents, contrary to the radio emissions simultaneous with the last TGF peak in multipeak TGFs (as well as with single TGFs).

## 6. Conclusions

### 6.1. RHESSI Clock Offsets and TGF-WWLLN Matches

RHESSI is one of the three currently operational satellites which can be used for the TGF observations. The off-line search algorithm developed by *Grefenstette et al.* [2009] and improved by *Gjesteland et al.* [2012], served to find a huge amount of TGFs. So far the RHESSI proved to be a very efficient TGF instrument. However, its clock offset has prevented the TGF community of the precise timing analysis which involves comparisons with the other types of data (WWLLN, etc). In the presented work we have estimated the RHESSI clock offset to an uncertainty level better than 100 μs, providing its values for the three time periods which cover the whole period of RHESSI observations.

We analyzed RHESSI TGF-WWLLN matches and relate the resulting distribution to the results obtained by the other groups [*Connaughton et al.*, 2013; *Marisaldi et al.*, 2015] for Fermi-GBM and AGILE satellites (see Figure 2). The resulting distribution has a remarkable similarity to the distributions obtained for Fermi GBM and AGILE TGF-WWLLN matches. Standard deviations of the three distributions are very close to each other, which implies that the resulting uncertainty could originate mainly from the WWLLN uncertainty and at the same time could refer to the natural variability of the process of radio emission produced by the TGFs.

The performed analysis of the RHESSI TGF-WWLLN matches allowed to evaluate the set of three distinct clock offsets related to the three observation periods of the RHESSI satellite. Those periods, offset values and their uncertainties are listed in Table 1.

The consistency of these results is independently confirmed by the RHESSI ground team and by the study of *Østgaard et al.* [2015] on the weak TGF search based on pileup statistical analysis of the RHESSI and WWLLN data (see Figure 3).

The evaluation of the RHESSI clock offsets opens a possibility of the combined analysis of RHESSI TGFs with any other data that involve precise timing (down to at least a 100 μs accuracy level).

### 6.2. Radio Emission From Multiple-Peak TGFs

We have found that for the multipeak TGFs the radio signal in all of the 16 revealed cases comes simultaneously with the last TGF peak. We could not find TGF-WWLLN pairs that contradict these observations. From the literature we found two Fermi GBM double TGFs with WWLLN detections simultaneous with the second TGF peaks and no contradicting cases [*Connaughton et al.*, 2010]. Analysis of the multipeak AGILE TGF-WWLLN matches performed by *Marisaldi et al.* [2014] revealed one multipeak TGF with WWLLN sferic simultaneous to the last TGF peak and one multipeak TGF with WWLLN detection simultaneous with the last peak was found by the authors. No contradicting cases were found. Also, one BATSE three-peak TGF was reported in association with three VLF sferics, and one double TGF in association with a single VLF sferic [*Cohen et al.*, 2006]. However, the source locations for those VLF sferics could not be determined.

We discussed two possible scenarios that could explain observed asymmetry between the multiple TGFs and radio emissions. The key point of one of the proposed mechanisms relates to the possibility that the TGFs could be generated along the axes tilted from vertical direction due to the tortuosity of the leader propagation path. However, the analysis of the VLF sferics associated with two double TGFs makes this scenario unlikely.

Another possible scenario attributes VLF emission to the recoil currents during the +IC leader stepping and the recoil current after the leader attachment (when the most powerful VLF sferic is emitted). In this scenario any VLF emission generated by the TGFs is assumed to be not strong enough and could be dominated by the +IC leader radiation. Observed asymmetry in multipeak TGF-WWLLN-VLF matching pattern could be due to the fact that the most favorable conditions for the TGF generation are produced by the leader when it develops to its full length, concentrating the maximal potential drop and the maximal local electric field strength in front of its tip, just before the attachment to a counterleader. After the attachment a recoil current wave produces VLF radio emission which masks any possible VLF radiation from the TGF itself. To discriminate between the TGF radiation and the RS-like radiation from the leader we propose for further studies to concentrate the efforts on the search and analysis of the radio emissions simultaneous with the first peak of the multipeak TGFs, when possible recoil currents from leader stepping are expected to be weaker than the final RS-like recoil current after the attachment, giving the possibility to distinguish between the leader processes radiation and VLF emission from the TGF producing electrons.

**Acknowledgments**

This study was supported by the European Research Council under the European Union's Seventh Framework Programme (FP7/2007-2013)/ERC grant agreement 320839 and the Research Council of Norway under contracts 208028/F50, 216872/F50, and 223252/F50 (CoE). We thank the RHESSI team for the use of RHESSI data and software. We thank the WWLLN team and the institutions contributing to WWLLN. All raw RHESSI data can be downloaded from: http://hesperia.gsfc.nasa.gov/ssw/hessi/doc/guides/hessi_data_access.htm. WWLLN data cannot be freely distributed and have to be purchased at http://wwlln.net. All other used data can be provided by the authors on request (Andrey.Mezentsev@uib.no).

## References

Boggs, S. E., A. Zoglauer, E. Bellm, K. Hurley, R. P. Lin, D. M. Smith, C. Wigger, and W. Hajdas (2007), The giant flare of 2004 December 27 from SGR 1806-20, *Astrophys. J.*, *661*, 458–467, doi:10.1086/516732.

Briggs, M. S., et al. (2010), First results on terrestrial gamma ray flashes from the Fermi Gamma-ray Burst Monitor, *J. Geophys. Res.*, *115*, A07323, doi:10.1029/2009JA015242.

Celestin, S., and V. P. Pasko (2011), Energy and fluxes of thermal runaway electrons produced by exponential growth of streamers during the stepping of lightning leaders and in transient luminous events, *J. Geophys. Res.*, *116*, A03315, doi:10.1029/2010JA016260.

Celestin, S., W. Xu, and V. P. Pasko (2012), Terrestrial gamma ray flashes with energies up to 100 MeV produced by nonequilibrium acceleration of electrons in lightning, *J. Geophys. Res.*, *117*, A05315, doi:10.1029/2012JA017535.

Cohen, M. B., U. S. Inan, and G. Fishman (2006), Terrestrial gamma ray flashes observed aboard the Compton Gamma Ray Observatory/Burst and Transient Source Experiment and ELF/VLF radio atmospherics, *J. Geophys. Res.*, *111*, D24109, doi:10.1029/2005JD006987.

Cohen, M. B., U. S. Inan, R. K. Said, and T. Gjesteland (2010), Geolocation of terrestrial gamma-ray flash source lightning, *Geophys. Res. Lett.*, *37*, L02801, doi:10.1029/2009GL041753.

Coleman, L. M., T. C. Marshall, M. Stolzenburg, T. Hamlin, P. R. Krehbiel, W. Rison, and R. J. Thomas (2003), Effects of charge and electrostatic potential on lightning propagation, *J. Geophys. Res.*, *108*(D9), 4298, doi:10.1029/2002JD002718.

Collier, A. B., T. Gjesteland, and N. Østgaard (2011), Assessing the power law distribution of TGFs, *J. Geophys. Res.*, *116*, A10320, doi:10.1029/2011JA016612.

Connaughton, V., et al. (2010), Associations between Fermi Gamma-ray Burst Monitor terrestrial gamma ray flashes and sferics from the World Wide Lightning Location Network, *J. Geophys. Res.*, *115*, A12307, doi:10.1029/2010JA015681.

Connaughton, V., et al. (2013), Radio signals from electron beams in terrestrial gamma ray flashes, *J. Geophys. Res. Space Physics*, *118*, 2313–2320, doi:10.1029/2012JA018288.

Cummer, S. A., Y. Zhai, W. Hu, D. M. Smith, L. I. Lopez, and M. A. Stanley (2005), Measurements and implications of the relationship between lightning and terrestrial gamma ray flashes, *Geophys. Res. Lett.*, *32*, L08811, doi:10.1029/2005GL022778.

Cummer, S. A., G. Lu, M. S. Briggs, V. Connaughton, G. J. Fishman, and J. R. Dwyer (2011), The lightning-TGF relationship on microsecond timescales, *Geophys. Res. Lett.*, *38*, L14810, doi:10.1029/2011GL048099.

Cummer, S. A., M. S. Briggs, J. R. Dwyer, S. Xiong, V. Connaughton, G. J. Fishman, G. Lu, F. Lyu, and R. Solanki (2014), The source altitude, electric current, and intrinsic brightness of terrestrial gamma ray flashes, *Geophys. Res. Lett.*, *41*, 8586–8593, doi:10.1002/2014GL062196.

Cummer, S. A., F. Lyu, M. S. Briggs, G. Fitzpatrick, O. J. Roberts, and J. R. Dwyer (2015), Lightning leader altitude progression in terrestrial gamma ray flashes, *Geophys. Res. Lett.*, *42*, 7792–7798, doi:10.1002/2015GL065228.

Dwyer, J. R. (2012), The relativistic feedback discharge model of terrestrial gamma ray flashes, *J. Geophys. Res.*, *117*, A02308, doi:10.1029/2011JA017160.

Dwyer, J. R., and S. A. Cummer (2013), Radio emissions from terrestrial gamma ray flashes, *J. Geophys. Res. Space Physics*, *118*, 3769–3790, doi:10.1002/jgra.50188.

Dwyer, J. R., and M. A. Uman (2014), The physics of lightning, *Phys. Rep.*, *534*, 147–241, doi:10.1016/j.physrep.2013.09.004.

Dwyer, J. R., B. W. Grefenstette, and D. M. Smith (2008), High-energy electron beams launched into space by thunderstorms, *Geophys. Res. Lett.*, *35*, L02815, doi:10.1029/2007GL032430.

Dwyer, J. R., D. M. Smith, and S. A. Cummer (2012), High-energy atmospheric physics: Terrestrial gamma-ray flashes and related phenomena, *Space Sci. Rev.*, *173*, 133–196, doi:10.1007/s11214-012-9894-0.

Fishman, G. J., et al. (1994), Discovery of intense gamma-ray flashes of atmospheric origin, *Science*, *264*, 1313–1316, doi:10.1126/science.264.5163.1313.

Foley, S., et al. (2014), Pulse properties of terrestrial gamma-ray flashes detected by the Fermi Gamma-ray Burst Monitor, *J. Geophys. Res. Space Physics*, *119*, 5931–5942, doi:10.1002/2014JA019805.

Gjesteland, T., N. Østgaard, P. H. Connell, J. Stadsnes, and G. J. Fishman (2010), Effects of dead time losses on terrestrial gamma ray flash measurements with the Burst and Transient Source Experiment, *J. Geophys. Res.*, *115*, A00E21, doi:10.1029/2009JA014578.

Gjesteland, T., N. Østgaard, A. B. Collier, B. E. Carlson, M. B. Cohen, and N. G. Lehtinen (2011), Confining the angular distribution of terrestrial gamma ray flash emission, *J. Geophys. Res.*, *116*, A11313, doi:10.1029/2011JA016716.

Gjesteland, T., N. Østgaard, A. B. Collier, B. E. Carlson, C. Eyles, and D. M. Smith (2012), A new method reveals more TGFs in the RHESSI data, *Geophys. Res. Lett.*, *39*, L05102, doi:10.1029/2012GL050899.

Gjesteland, T., N. Østgaard, S. Laviola, M. M. Miglietta, E. Arnone, M. Marisaldi, F. Fuschino, A. B. Collier, F. Fabro, and J. Montanya (2015), Observation of intrinsically bright terrestrial gamma ray flashes from the Mediterranean basin, *J. Geophys. Res. Atmos.*, *120*, 12,143–12,156, doi:10.1002/2015JD023704.

Grefenstette, B. W., D. M. Smith, B. J. Hazelton, and L. I. Lopez (2009), First RHESSI terrestrial gamma ray flash catalog, *J. Geophys. Res.*, *114*, A02314, doi:10.1029/2008JD013721.

Hazelton, B. J., B. W. Grefenstette, D. M. Smith, J. R. Dwyer, X.-M. Shao, S. A. Cummer, T. Chronis, E. H. Lay, and R. H. Holzworth (2009), Spectral dependence of terrestrial gamma-ray flashes on source distance, *Geophys. Res. Lett.*, *36*, L01108, doi:10.1029/2008GL035906.

Hutchins, M. L., R. H. Holzworth, J. B. Brundell, and C. J. Rodger (2012), Relative detection efficiency of the World Wide Lightning Location Network, *Radio Sci.*, *47*, RS6005, doi:10.1029/2012RS005049.

Inan, U. S., S. C. Reising, G. J. Fishman, and J. M. Horack (1996), On the association of terrestrial gamma-ray bursts with lightning and implications for sprites, *Geophys. Res. Lett.*, *23*(9), 1017–1020.

Lu, G., R. J. Blakeslee, J. Li, D. M. Smith, X.-M. Shao, E. W. McCaul, D. E. Buechler, H. J. Christian, J. M. Hall, and S. A. Cummer (2010), Lightning mapping observation of a terrestrial gamma-ray flash, *Geophys. Res. Lett.*, *37*, L11806, doi:10.1029/2010GL043494.

Lu, G., S. A. Cummer, J. Li, F. Han, D. M. Smith, and B. W. Grefenstette (2011), Characteristics of broadband lightning emissions associated with terrestrial gamma ray flashes, *J. Geophys. Res.*, *116*, A03316, doi:10.1029/2010JA016141.

Lu, G., et al. (2013), Coordinated observations of sprites and in-cloud lightning flash structure, *J. Geophys. Res. Atmos.*, *118*, 6607–6632, doi:10.1002/jgrd.50459.

Marisaldi, M., et al. (2014), Properties of terrestrial gamma ray flashes detected by AGILE MCAL below 30 MeV, *J. Geophys. Res. Space Physics*, *119*, 1337–1355, doi:10.1002/2013JA019301.

Marisaldi, M., et al. (2015), Enhanced detection of terrestrial gamma-ray flashes by AGILE, *Geophys. Res. Lett.*, *42*, 9481–9487, doi:10.1002/2015GL066100.

Marshall, T., M. Stolzenburg, S. Karunarathne, S. A. Cummer, G. Lu, H.-D. Betz, M. S. Briggs, V. Connaughton, and S. Xiong (2013), Initial breakdown pulses in intracloud lightning flashes and their relation to terrestrial gamma ray flashes, *J. Geophys. Res. Atmos.*, *118*, 10,907–10,925, doi:10.1002/jgrd.50866.

Nemiroff, R. J., J. T. Bonnell, and J. P. Norris (1997), Temporal and spectral characteristics of terrestrial gamma flashes, *J. Geophys. Res.*, *102*, 9659–9665.

Østgaard, N., T. Gjesteland, B. E. Carlson, A. B. Collier, S. A. Cummer, G. Lu, and H. J. Christian (2013), Simultaneous observations of optical lightning and terrestrial gamma ray flash from space, *Geophys. Res. Lett.*, *40*, 2423–2426, doi:10.1002/grl.50466.

Østgaard, N., K. H. Albrechtsen, T. Gjesteland, and A. B. Collier (2015), A new population of terrestrial gamma-ray flashes in the RHESSI data, *Geophys. Res. Lett.*, *42*, 10,937–10,942, doi:10.1002/2015GL067064.

Palmer, D. M., et al. (2005), A giant gamma-ray flare from the magnetar SGR 1806-20, *Nature*, *434*, 1107–1109, doi:10.1038/nature03525.

Rison, W., P. R. Krehbiel, M. G. Stock, H. E. Edens, X.-M. Shao, R. J. Thomas, M. A. Stanley, and Y. Zhang (2016), Observations of narrow bipolar events reveal how lightning is initiated in thunderstorms, *Nat. Commun.*, *7*, 10721, doi:10.1038/ncomms10721.

Rodger, C. J., J. B. Brundell, and R. L. Dowden (2005), Location accuracy of VLF World-Wide Lightning Location (WWLL) network: Post-algorithm upgrade, *Ann. Geophys.*, *23*, 277–290.

Rodger, C. J., S. Werner, J. B. Brundell, E. H. Lay, N. R. Thomson, R. H. Holzworth, and R. L. Dowden (2006), Detection efficiency of the VLF World-Wide Lightning Location Network (WWLLN): Initial case study, *Ann. Geophys.*, *24*, 3197–3214.

Shao, X.-M., T. Hamlin, and D. M. Smith (2010), A closer examination of terrestrial gamma-ray flash-related lightning processes, *J. Geophys. Res.*, *115*, A00E30, doi:10.1029/2009JA014835.

Smith, D. M., et al. (2002), The RHESSI spectrometer, *Sol. Phys.*, *210*, 33–60.

Stanley, M. A., X.-M. Shao, D. M. Smith, L. I. Lopez, M. B. Pongratz, J. D. Harlin, M. Stock, and A. Regan (2006), A link between terrestrial gamma-ray flashes and intracloud lightning discharges, *Geophys. Res. Lett.*, *33*, L06803, doi:10.1029/2005GL025537.

van der Velde, O. A., A. Mika, S. Soula, C. Haldoupis, T. Neubert, and U. S. Inan (2006), Observations of the relationship between sprite morphology and in-cloud lightning processes, *J. Geophys. Res.*, *111*, D15203, doi:10.1029/2005JD006879.

van der Velde, O. A., J. Montanya, S. Soula, N. Pineda, and J. Mlynarczyk (2014), Bidirectional leader development in sprite-producing positive cloud-to-ground flashes: Origins and characteristics of positive and negative leaders, *J. Geophys. Res. Atmos.*, *119*, 12,755–12,779, doi:10.1002/2013JD021291.

Winn, W. P., G. D. Aulich, S. J. Hunyady, K. B. Eack, H. E. Edens, P. R. Krehbiel, W. Rison, and R. G. Sonnenfeld (2011), Lightning leader stepping, K changes, and other observations near an intracloud flash, *J. Geophys. Res.*, *116*, D23115, doi:10.1029/2011JD015998.